\documentstyle[12pt]{article}
\begin{document}
\title{Inelastic Collision and Switching of\\
 Coupled Bright Solitons  in  Optical Fibers}
\author { \\R. Radhakrishnan${}^1$, M. Lakshmanan${}^1$,\\
and\\ J. Hietarinta${}^2$\\
$^1$Centre for Nonlinear Dynamics, Department of Physics,
\\ Bharathidasan University,
Tiruchirapalli - 620 024, India\\
$^2$Department of Physics, University of Turku\\ FIN-20014,
Turku, Finland (email: hietarin@utu.fi)}
\maketitle
\begin{abstract}
By constructing the general six-parameter bright two-soliton solution
of the integrable coupled nonlinear Schr\"{o}dinger equation (Manakov
model) using the Hirota method, we find that the solitons exhibit
certain novel inelastic collision properties, which have not been
observed in any other (1+1) dimensional soliton system so far.  In
particular, we identify the exciting possibility of switching solitons
between modes by changing the phase.  However, the standard elastic
collision property of solitons is regained with specific choices of
parameters.
\end{abstract}


\newpage
Recent developments in the field of optical solitons in fibers have
already revealed the possibility of overcoming the limitations on the
speed and distance of linear-wave transmission systems \cite{R1,R3}.
However, the interaction between optical solitons influences directly
the capacity and quality of communication \cite{R3,R5}. Since the
nonlinear Schr\"{o}dinger (NLS) equation is used as a mathematical
model for such studies, considerable attention has been devoted to
analyze the nature of collision between NLS solitons \cite{R1,R3,R5,R7}. In
more general physical situations \cite{R9}, coupled NLS equations are
often used to describe the interaction among the modes in nonlinear
optics, as for example, in the case of birefringent \cite{R11} and
other two-mode \cite{R12} fibers. Here we consider the integrable
coupled NLS equation of Manakov type \cite{R13},
\begin{equation}
\begin{array}{rcl}
iq_{1x}+q_{1tt}+2\mu\left (\left | q_1 \right |^{2}+\left |
q_2\right |^{2}\right ) q_1 &=& 0,\\ 
iq_{2x}+q_{2tt}+2\mu\left (\left | q_1 \right |^{2}+\left |
q_2\right |^{2}\right ) q_2 &=& 0,
\end{array}\label{E:1}
\end{equation} 
where $q_1$ and $q_2$ are slowly varying envelopes of the two
interacting optical modes, the variables ${\it x}$ and ${\it t}$ are
the normalized distance and time, and $\mu$ is a positive parameter.

Exact soliton solutions have been derived \cite{R13,R14,R15,R16} for
the system (\ref{E:1}) with different procedures. In \cite{R14}, using
bright one-soliton solutions of the system (\ref{E:1}), many physical
phenomena such as birefringence property, soliton trapping and
daughter wave ('shadow') formation are studied. Further, as noted in
\cite{R17}, the Manakov model (\ref{E:1}) is important in describing
the effects of averaged random birefringence on an orthogonally
polarized pulse in a real fiber. In addition, by considering the
analytic solution of the system (\ref{E:1}), conditions have been
established for soliton switching and energy coupling among the two
modes in a nonlinear fiber \cite{R19}. Recently in \cite{R15}, two of
the present authors (R.R and M.L) have derived bright and dark
multi-soliton solutions of the system (\ref{E:1}). The interaction
between bright and dark solitary waves is studied in \cite{R20}.

In this paper, we focus our attention on the system (\ref{E:1}) to
derive a more general bright two-soliton solution than the ones that
have been presented in the literature \cite{R15} by using the Hirota
method \cite{R21,R22}.  The asymptotic behavior of this solution is
studied further in order to explain the collision properties of
coupled solitons in the anomalous dispersion region.

It is often stated in the literature \cite{R13,R23} that the
two-soliton solution is very difficult to derive and too complicated
to analyze. [Nevertheless Manakov \cite{R13} was able to make an
asymptotic analysis of the solitons.]  In this paper we present for
the first time the completely general two-soliton solution to
(\ref{E:1}) in explicit form, which can be readily used asymptotic
analysis, numerical computations, etc. Our explicit solution not only
clarifies Manakov's observation on the asymptotic behavior but also
helps to realize the exciting possibility of a novel type of inelastic
collision allowing switching between components via phase change; this
in contrast to the standard elastic collision usually observed in
(1+1)-dimensional soliton systems. [However, the standard elastic
collision property of the solution is recovered when restrictions are
imposed on some of the free parameters.] The situation is reminiscent
of the dromion solutions in (2+1)-dimensional systems such as the
Davey-Stewartson equation \cite{R24,R25} where inelasticity has been
observed \cite{R25,R26} in the scattering process. The details of the
present study are as follows.

Recently, a special form of the bright two-soliton solution for
Eq.(\ref{E:1}) with five arbitrary complex parameters has been derived
in \cite{R15}, using the Hirota method \cite{R21,R22}. The first step
in this method is to transform the system (\ref{E:1}) into the Hirota
bilinear form. For this purpose we use the transformations \cite{R15},
$q_1 = g/f$ and $ q_2 = h/f,$ with $f$ real, to obtain the bilinear
form of (\ref{E:1}) as
\begin{equation}
\left ( iD_x+D_t^2\right )g\cdot f = 0,\quad
\left ( iD_x+D_t^2\right )h\cdot f = 0,\quad
D_t^2f\cdot f = 2\mu\left ( gg^*+hh^*\right ).
\label{E:2}\end{equation}
Here the operator $D$ is defined by $D_x^n\, a\cdot
b=(\partial_{x_1}-\partial_{x_2})a(x_1)b(x_2)|_{x=x_1=x_2}$.  The
one-soliton solution to (\ref{E:2}) is given by $f=\alpha
e^{\eta_1},\, h=\beta e^{\eta_1},\, g=1+e^{\eta_1+\eta_1^*+R}$, that
is,
\begin{equation}
q_1=\frac{\alpha e^{\eta_1}}{1+e^{\eta_1+\eta_1^*+R}},\quad 
q_2=\frac{\beta e^{\eta_1}}{1+e^{\eta_1+\eta_1^*+R}},\quad
 e^R=\frac{\mu(|\alpha|^2+|\beta|^2)}{(k_1+k_1^*)^2}
\label{E:1ss}
\end{equation}
where
\begin{equation}
\eta_j = k_j\left ( t+ik_jx\right )+\eta_j^{(0)}.
\end{equation}
The arbitrary complex parameters in (\ref{E:1ss}) are $k_1$ and any
two of the set $\alpha, \,\beta,\, \eta_1^{(0)}$.  This solution can
also be written in the more conventional form
\begin{equation}
q_i=\frac{k_{1R}\,A_i\,e^{i\eta_{1I}}}{\cosh(\eta_{1R}+\phi)},
\label{E:ss}
\end{equation}
where 
\[\phi=\frac12R,\quad 
A_1=\frac{\alpha}\Delta,\quad A_2=\frac{\beta}\Delta,\quad
\Delta={\sqrt{\mu(|\alpha|^2+|\beta|^2)}}.
\]
and we have introduced the subscripts $R$ and $I$ for the real and
imaginary parts of the quantity in question. (A positive root has been
used to define $e^{\frac12 R}$ and hence $\phi$ is real.) From this
form it is easy to identify the amplitude $A$ and the phase $\phi$.
We also note that $k_{1R}$ and $k_{1I}$ determine the amplitude and
velocity of solitons, and in Manakov's notation \cite{R13} $A_i$
corresponds to the unit polarization vector of the soliton.

The two-soliton solution can be obtained by substituting into
(\ref{E:2}) the expansion $g = \chi g_1+\chi^3 g_3,\, h = \chi
h_1+\chi^3 h_3,$ and $f = 1+\chi^2 f_2+\chi^4 f_4$, where $\chi$ is
the formal expansion parameter. The main problem is to choose the
proper ansatz for $g_1,\,h_1$. In \cite{R15} a bright two-soliton
solution was derived in this way assuming the input expression $g_1 =
e^{\eta_1}+e^{\eta_2},\, h_1 = e^\varepsilon\left (
e^{\eta_1}+e^{\eta_2} \right )$. The resulting two-soliton solution
shows the standard elastic collision with a phase-shift.

However, for any integrable equation it must be possible to combine
{\em any} pair of one-soliton solutions into a two-soliton solution
\cite{R22}, and therefore it should also be possible to start with
$g_1$ and $h_1$ given by
\begin{equation}
g_1 = \alpha_1e^{\eta_1}+\alpha_2e^{\eta_2},\quad
h_1 = \beta_1e^{\eta_1}+\beta_2e^{\eta_2},
\end{equation}
and this way generate a more general bright two-soliton solution with
{\em six} arbitrary complex parameters $k_1$, $k_2$, $\alpha_1$,
$\alpha_2$, $\beta_1$, and $\beta_2$ (Note that the parameters
$\eta_j^ {(0)}$ in (4) have been absorbed into $\alpha_j$ and
$\beta_j$). This is indeed possible.  By following the equations
(10-28) in \cite{R15}, we obtain the most general expressions for the
two optical modes $q_1$ and $q_2$ as
\begin{eqnarray}
q_1 &=& \frac{\alpha_1e^{\eta_1}+\alpha_2e^{\eta_2}+
e^{\eta_1+\eta_1^*+\eta_2+\delta_1}+
e^{\eta_1+\eta_2+\eta_2^*+\delta_2}}{1+e^{\eta_1+\eta_1^*+R_1}+
e^{\eta_1+\eta_2^*+\delta_0}+e^{\eta_1^*+\eta_2+\delta_0^*}+
e^{\eta_2+\eta_2^*+R_2}+e^{\eta_1+\eta_1^*+\eta_2+
\eta_2^*+R_3}},\nonumber\\ & & \label{E:5}
\\
q_2 &=& \frac{\beta_1e^{\eta_1}+\beta_2e^{\eta_2}+
e^{\eta_1+\eta_1^*+\eta_2+\delta_{1}'}+e^{\eta_1+\eta_2+\eta_2^*+
\delta_{2}'}}{1+e^{\eta_1+\eta_1^*+R_1}+e^{\eta_1+\eta_2^*+\delta_0}+
e^{\eta_1^*+\eta_2+\delta_0^*}+e^{\eta_2+\eta_2^*+R_2}+
e^{\eta_1+\eta_1^*+\eta_2+\eta_2^*+R_3}},\nonumber
\end{eqnarray}
where [note that $R_i$ are real]
\begin{eqnarray}
e^{\delta_0}&=&\frac{\kappa_{12}}{k_1+k_2^*},\quad
e^{R_1}=\frac{\kappa_{11}}{k_1+k_1^*},\quad
e^{R_2}=\frac{\kappa_{22}}{k_2+k_2^*},\\
e^{\delta_1}&=&\frac{k_1-k_2}{(k_1+k_1^*)(k_1^*+k_2)}
 (\alpha_1\kappa_{21}-\alpha_2\kappa_{11}),\\
e^{\delta_2}&=&\frac{k_2-k_1}{(k_2+k_2^*)(k_1+k_2^*)}
 (\alpha_2\kappa_{12}-\alpha_1\kappa_{22}),\\
e^{\delta'_1}&=&\frac{k_1-k_2}{(k_1+k_1^*)(k_1^*+k_2)}
 (\beta_1\kappa_{21}-\beta_2\kappa_{11}),\\
e^{\delta'_2}&=&\frac{k_2-k_1}{(k_2+k_2^*)(k_1+k_2^*)}
 (\beta_2\kappa_{12}-\beta_1\kappa_{22}),\\
e^{R_3}&=&\frac{|k_1-k_2|^2}{(k_1+k_1^*)(k_2+k_2^*)|k_1+k_2^*|^2}
(\kappa_{11}\kappa_{22}-\kappa_{12}\kappa_{21}),\\
\mbox{ and }\quad\kappa_{ij}&=& \frac{\mu(
\alpha_i\alpha_j^*+\beta_i\beta_j^*)}{k_i+k_j^*}.
\end{eqnarray}

Does the introduction of the additional parameters make any
qualitative change in the behavior of soliton? The answer is yes and
we find the novel result that the above general solution (\ref{E:5})
corresponds to an {\em inelastic} collision of two bright solitons, as
long as $\alpha_1:\alpha_2 \neq \beta_1:\beta_2$. In order to see
this, we analyze the asymptotic form of the solution (\ref{E:5}).

One important advantage of Hirota's method is that the solution allows
easy analysis of the asymptotic behavior by taking a given
$\eta_{iR}\to\pm\infty$ in (\ref{E:5}) and comparing the result with
the one-soliton solution (\ref{E:1ss}). The interpretation of the
result in terms of the actual motion of the solitons depends on the
signs of $k_{iR}$ and $k_{iI}$. In general
$\eta_{iR}=k_{iR}(t-2k_{iI}x)$ so that soliton $j$ is located in the
vicinity of the line $t=2k_{jI}x$.  Let us change to the frame
comoving with soliton $j$ (coordinatized by $\xi$) by putting
$x=(t-\xi)/(2k_{jI})$. Then $\eta_{jR}=k_{jR}\xi$, while for the other
soliton $m$ we get $\eta_{mR}= k_{mR}(1- \frac{k_{mI}}{k_{jI}})t
+k_{mR}\frac{k_{mI}}{k_{jI}}\xi$.  Thus if we have, for example,
$k_{iR}>0$ and $k_{1I}k_{2I}<0$, which corresponds to a head-on
collision, we find that $\eta_{iR}\to\pm\infty$ corresponds to
$t\to\pm\infty$. [Correspondence between the signs of $x\to\pm\infty$
and $\eta_{iR}\to\pm\infty$ is obtained in a similar way and leads to
a slightly different dependence on the signs of the $k$'s.]

In each limit $\eta_{iR}\to\pm\infty$ the resulting $q_i$ can be
written as in (\ref{E:ss}) with different amplitudes $A$ and phases
$\phi$. Let us denote by $A_i^{k\pm}$ the amplitude of the component
$q_i$ of the soliton $k$ as the other soliton goes to $\pm\infty$, and
similarly for the phase $\phi$. Furthermore let us define the phase
shift $\Phi^k=\phi^{k+}-\phi^{k-}$, and a ``transition matrix''
$T_i^k$ by $A_i^{k+}=A_i^{k-}\,T_i^k$. [Since the magnitudes of $A$
and $T$ are the most interesting quantities we will not study their
phases here.]  We find the following results:

\noindent
Soliton 1:
\begin{eqnarray}
|A_1^{1-}|&=&|\alpha_1|/\Delta_1,\quad 
|A_2^{1-}|=|\beta_1|/\Delta_1,\\
|T_1^1|&=&\frac{|1-\lambda_2\alpha_2/\alpha_1|}
{\sqrt{|1-\lambda_1\lambda_2|}},\quad
|T_2^1|=\frac{|1-\lambda_2\beta_2/\beta_1|}
{\sqrt{|1-\lambda_1\lambda_2|}},\\
\phi^{1-}&=&\frac12R_1,\quad \Phi^1=\frac12(R_3-R_1-R_2).
\end{eqnarray}

\noindent
Soliton 2:
\begin{eqnarray}
|A_1^{2-}|&=&|\alpha_2|/\Delta_2,\quad
|A_2^{2-}|=|\beta_2|/\Delta_2,\\
|T_1^2|&=&\frac{|1-\lambda_1\alpha_1/\alpha_2|}
{\sqrt{|1-\lambda_1\lambda_2|}}\quad
|T_2^2|=\frac{|1-\lambda_1\beta_1/\beta_2|}
{\sqrt{|1-\lambda_1\lambda_2|}},\\
\phi^{2-}&=&\frac12R_2,\quad\Phi^2=\frac12(R_3-R_1-R_2)(=\Phi^1).
\end{eqnarray}
where
\begin{equation}
\Delta_i=\sqrt{\mu(|\alpha_i|^2+|\beta_i|^2)},\quad
\lambda_1=\kappa_{21}/\kappa_{11},\quad
\lambda_2=\kappa_{12}/\kappa_{22}.
\end{equation}
The expressions above are scale invariant in the way that, for
example, allows us to take $\beta_i=1$. If we also have
$\alpha_1=\alpha_2$ then one can easily verify that $|T_i^j|=1$, which
implies perfect elastic scattering.  We also note that the
noninteracting stationary pulses discussed in \cite{R23} follow with
the choice $\alpha_2=\beta_1=0, k_{1I}=k_{2I}=0$.

One can verify that $|A_1^{i-}|^2+|A_2^{i-}|^2= |A_1^{i+}|^2+
|A_2^{i+}|^2 =1/\mu$ so that the {\em total} intensity of each soliton
is conserved. However, the {\em distribution} of this intensity among
the two component fields can change during collision.  It turns out
that an inelastic effect can be obtained just by changing the
relative {\em phases} of the parameters $\alpha_i, \beta_i$.

In figure 1 we have a picture of a head-on collision with
$k_1=1+i,\,k_2=2-i$, $\beta_i=1,\,\alpha_1=1$ and
$\alpha_2=(39+i80)/89$. We still have $|\alpha_2|=1$, but its
nontrivial phase ($\alpha_2=e^{i\theta},\,\theta\approx 64^o$) is
enough to cause quite dramatic non-elasticity. The initial
time-profiles at both ends of the $x$-axis are evenly split between
the two components, but the profile observed later at the large
positive $x$ end is almost completely in component 2.  If we had
chosen $\alpha_2=1$ the scattering would have been elastic.  Thus, by
changing the relative phase of just one component of one soliton we
are able to change the final state quite dramatically.

Finally, we wish to connect explicitly our results on inelastic
collision with that of the work of Manakov in [8].  In formulating the
results that he obtained by an asymptotic analysis of the inverse
scattering problem associated with (1), Manakov used a (complex unit)
polarization vector obtained from the amplitudes of the solitons ((19)
and (20) in [8]). Correspondingly, in our case the initial polarization
vectors are
\begin{eqnarray}
\hat c_i = (\alpha_i,\beta_i)^T/\sqrt{|\alpha_i|^2+|\beta_i|^2},
\end{eqnarray}
while the final vectors can be expressed as
\begin{eqnarray}
\hat c_1{}' &=& \frac1\chi\frac{k_1-k_2}{k_1+k_2^*}
\left(\hat c_1-\frac{k_2+k_2^*}{k_1+k_2^*}(\hat c_1\cdot \hat c_2{}^*)
\hat c_2\right),\\ 
\hat c_2{}' &=& \frac1\chi\frac{k_2-k_1}{k_2+k_1^*}
\left(\hat c_2-\frac{k_1+k_1^*}{k_2+k_1^*}(\hat c_2\cdot \hat c_1{}^*)
\hat c_1\right), \\
\mbox{where }\qquad\chi&=&\frac{|k_1-k_2|}{|k_1+k_2^*|}\left\{
1-\frac{(k_1+k_1^*)(k_2+k_2^*)}
{|k_1+k_2^*|^2}|\hat c_1\cdot \hat c_2{}^*|^2\right\}^{\frac12}.
\end{eqnarray}
(The connection between Manakov's $\zeta$ and our $k$ is
$\zeta_1=(-ik_2)^*,\, \zeta_2=-ik_2$.)

Manakov pointed out that during soliton collision their velocities and
amplitudes (intensities) do not change but the associated unit
polarization vectors do change provided they are neither parallel nor
orthogonal. Our observation, illustrated in Fig. 1, is that even if we
keep the initial {\it amplitudes} unchanged and just change their
relative {\it phases} we can change the amplitude distribution in the
final polarization vector.  The parametric choice associated with
Fig.1 is $ \hat c_1 = (1/\sqrt 2, 1/\sqrt 2)^T, \hat c_2 =
(-(39+80i)/(89\sqrt 2), 1/\sqrt 2)^T, \zeta_1 = 1+i, \zeta_2 =
-1-2i$. Each component here has magnitude $1/\sqrt2$.  After collision
the unit polarization vectors take the values $\hat
c_1{}'=(-0.0190+0.0593 i,-0.734+0.677 i), \hat c_2{}'=(0.422+0.272
i,0.462 - 0.731 i)$, with magnitude distribution $(0.0623,0.998)$ and
$ (0.502,0.865)$, respectively.  One can also note that the magnitude
of the first component of $\hat c_1{}'$ can be even made exactly zero if
we can also change the parameters $k_i$ suitably. However, in
practical applications it is probably only the relative phase that can
be easily modified.

To conclude, the general two-soliton solution (\ref{E:5}) of the
(1+1)-dimensional system (\ref{E:1}) exhibits a novel type of
inelastic collision, not seen in any other (1+1)-dimensional evolution
equation.  Naturally such a property will have important ramifications
in optical fiber communication such as providing intensity pump
sources, soliton switching and so on, which remain to be explored.

It will also be of interest to investigate the ramifications of the above
type of inelastic collision in non-integrable cases, for example
when the nonlinear cross coupling coefficients are different from one.

The work of R.R. and M.L. forms part of a Department of Science and
Technology research project. The work of J.H. is partially supported
by the Academy of Finland, project 31445.
 
\vskip 0.5cm
\parindent 0 cm
\subsection * { Figure Captions}
{Fig.1} { Intensity profiles $\left | q_1\right |$ (top) and $\left
| q_2\right |$ (bottom) of the head-on collision solution
(\ref{E:5}) with the parameter values $k_1=1+i,\,k_2=2-i$,
$\beta_1=\beta_2=\alpha_1=1$ and $\alpha_2=(39+i80)/89$.

\vskip 0.5cm
\parskip 0pt
\baselineskip 0pt

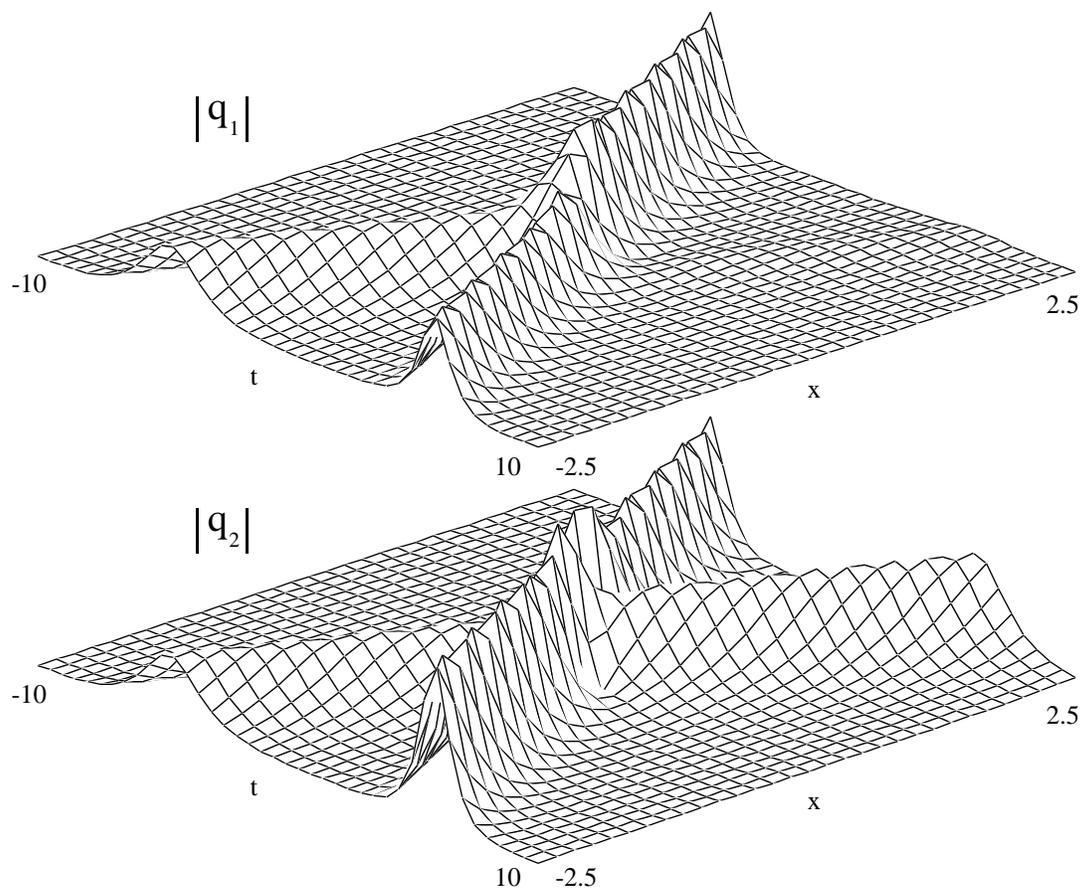
\end{document}